# The Impact of Triangular-Toothed Gears on the Functionality of the Antikythera Mechanism

**Authors:**

Prof. Esteban Guillermo Szigety

Facultad de Ingeniería - Universidad Nacional de Mar del Plata – Av. Juan B. Justo 4302 - Mar del Plata - Argentina

Correo de contacto: esteszige@gmail.com

Dr. Gustavo Francisco Arenas

CONICET- Facultad de Ingeniería - Universidad Nacional de Mar del Plata – Av. Juan B. Justo 4302  - Mar del Plata – Argentina

Correo de contacto: garenas@fi.mdp.edu.ar

**Abstract:**

The Antikythera Mechanism is based on a complex system of interconnected gears. Recent analyses have highlighted the influence of triangular tooth profiles and manufacturing inaccuracies on its performance. This study combines Alan Thorndike's analytical solution for the non-uniform motion caused by triangular teeth with Mike Edmunds' error model accounting for manufacturing imprecisions. We developed a computational program to simulate the behavior of the mechanism's pointers, integrating variables from both models. Since the impact of these variables is speculative, our results must be interpreted with caution. Our findings indicate that while the triangular shape of the teeth alone produces negligible errors, manufacturing inaccuracies significantly increase the likelihood of gear jamming or disengagement. Under our assumptions, the errors identified by Edmunds exceed the tolerable limits required to prevent failures. Consequently, either the mechanism never functioned or its actual errors were smaller than those reported by Edmunds. Although it seems unlikely that someone would build such a complex yet non-functional device, there are strong reasons to question whether Edmunds' values accurately reflect the mechanism's original errors.

### 1- Introduction general

The Antikythera Mechanism is a mechanical astronomical instrument discovered in an ancient shipwreck in the early 20th century. While the shipwreck has been dated to the decades around 60 BCE, there is no consensus on whether the mechanism was built shortly before the ship sank or much earlier. After spending twenty centuries underwater, the mechanism is incomplete and fragmented. However, the remaining pieces are sufficient to reconstruct its structure and main functions. The device featured several interlinked indicators, driven by a system of gears, which displayed the positions of the Moon and the Sun (and likely the planets) within the zodiac, the date according to the Egyptian calendar, and a Greek lunisolar calendar, as well as details of upcoming solar and lunar eclipses.[1]

The indicators are driven by a gear train, which introduces two main sources of error that affect their accuracy. First, the triangular shape of the teeth[2] results in non-uniform motion, causing acceleration and deceleration as each tooth engages. Second, the inevitable inaccuracies in gear manufacturing can also generate errors. Both sources of error have been previously analyzed by various scholars.

---

[1] For a systematic introduction to the research on the mechanism as well as an updated bibliography see Jones 2017.

[2] It seems that Derek De Solla Price was the first to inform about the triangular shape of the gears. Cfr. De Solla Price 1974.

Alan Thorndike (2019) developed an analytical solution to calculate the non-uniform motion produced by two interconnected gears with triangular teeth, like those of the Antikythera Mechanism. His main conclusion is that the ratio between the fastest and slowest speeds generated by the triangular teeth depends on the gear radii and the height of the triangular tooth.

Mike Edmunds calculated the deviation of the Antikythera Mechanism's main indicators from their theoretical values, caused by the non-ideal arrangement of the teeth, which exhibit both random and systematic positioning errors. He developed a model to study error propagation by first generating gears with ideally positioned teeth and then introducing random errors based on a normal distribution, along with a sinusoidal error of specified amplitudes, which he defined as systematic error. He examined manufacturing inaccuracies by measuring the displacement of the teeth and the center of the gears in CT scans of the fragments. After running his model with these input values, he concluded that the positions of the Saros and Metonic indicators were sufficiently accurate to display their intended values, but this was not the case for the lunar indicator. In particular, he noted that the error would obscure the anomalistic effect that the pin-and-slot device was designed to represent. He ultimately concluded that the primary purpose of the Antikythera Mechanism was more of an educational display than a tool for making practical and precise astronomical predictions.

Alan Thorndike's analysis provides an excellent foundation, but it is extremely complex to apply analytically to intricate gear trains like those of the Antikythera Mechanism, which involve multiple interconnected gear pairs. Moreover, Thorndike's analysis does not account for manufacturing errors, whether random and/or systematic errors in the distribution of the teeth. In contrast, Mike Edmunds' model does not consider the impact of errors due to the triangular shape of the teeth. His geometric analysis treats each gear as if it were a radius extending from the center, ignoring the actual shape of the gear—an approach that is appropriate for the analysis he conducted.

To achieve a more realistic analysis of the possible errors in the mechanism, we propose combining both approaches. We developed a computational model inspired by Thorndike's work to simulate the behavior of triangular teeth, allowing us to analyze the errors introduced by their shape. Modeling triangular-toothed gears without manufacturing errors resulted in negligible deviations in the positions of the indicators. We then incorporated Edmunds' errors into our model to predict their effects on the triangular teeth. Before doing so, it was necessary to formulate hypotheses regarding the proportion of certain variables, as will be explained later. These assumptions are essential because neither model independently requires precise definitions, but they become necessary when combined into a single model. The behavior of the final pointers, compared to theoretical results, showed no significant differences from the pointer deviations obtained by Edmunds when triangular-toothed gears were combined with manufacturing errors.

However, our model revealed numerous instances of gear jamming and disengagement caused by both the random and systematic distribution of the teeth. Since the Antikythera Mechanism has a single input of motion and all gears are interconnected, the jamming of any gear pair would bring the entire mechanism to a halt. Disengagement, on the other hand, would at least partially stop the indicators dependent on the disengaged pair, causing synchronization errors among the pointers.

As expected, jamming and decoupling depend significantly on the spacing between the gears. However, even with an optimized spacing, our analysis suggests that the gears of the MA would have needed to be manufactured with greater precision than what Edmunds observed in the computed tomography scans. According to our results, if the Antikythera Mechanism had the manufacturing errors measured by Edmunds, it would likely jam or stop before the solar indicator could complete a motion equivalent to 120 days (4 months).

In the second section, we analyze the analytical aspects of motion transmission in triangular-toothed gears. In the third section, we describe the model we developed to simulate the errors produced by perfect triangular gears (without manufacturing errors) and discuss the results. In the fourth section, we explain the incorporation of Edmunds' manufacturing errors into our model and examine the outcomes. In the fifth section, we explore in detail the impact of these errors on gear jamming and disengagement.

## 2- Concepts of triangular-profile gears

When two gears mesh, the ratio of their rotation periods corresponds to the ratio of their respective numbers of teeth. This relationship can be expressed as follows: if $\phi$ denotes the angular position of the driving gear with $N_1$ teeth, and $\theta$ represents the angular position of the driven gear with $N_2$ teeth, then:

$$\frac{\theta}{\phi} = \frac{N_1}{N_2} \quad [1]$$

The ratio $N_1/N_2$ is known as the gear ratio. Expression [1] becomes meaningful if we define an angular reference system in which the positive direction is adopted for both rotations, even though the gears rotate in opposite directions—one clockwise and the other counterclockwise. Additionally, it is assumed that both gears start from an initial angular position of 0°.

Equation [1] applies accurately for any $\phi$ that is a multiple of its angular pitch $360/N_1$, meaning whenever the driving gear rotates by an angle corresponding to an integer number of teeth. However, when applied to intermediate positions, Equation [1] is only an approximation, as the ratio $\theta/\phi$ depends on the shape of the gear tooth profile. The involute profile of modern gears ensures that Equation [1] holds for any $\phi$, implying that if the driving gear rotates at a constant angular velocity, the driven gear does as well. In contrast, triangular teeth cause acceleration and deceleration in the angular velocity of the driven gear, even when the driving gear moves uniformly.

Based on the work of Alan Thorndike, we have developed a complete analytical solution to determine the angular position $\theta$ from any angle $\phi$ in the case of two gears with triangular teeth. This solution, tested in Appendix I, is useful for analyzing transmission between two simple gears, although it presents limitations for trains of multiple interconnected gears due to the need to compose several functions. The variables involved in this solution are based on specific geometric parameters of triangular gears.

Figure 1 illustrates the concept of angular pitch, calculated as $360°/N$, where $N$ is the number of teeth on the gear. Additionally, two characteristic radii are defined: the outer radius $R$ and the inner radius $r$, from which the tooth height is derived as $h = R - r$. The distance $D$ represents the separation between the centers of the gears, while $G$ denotes the distance between the tip of a tooth on one gear and the valley of the opposing tooth when they are aligned, as shown in Figure 1.

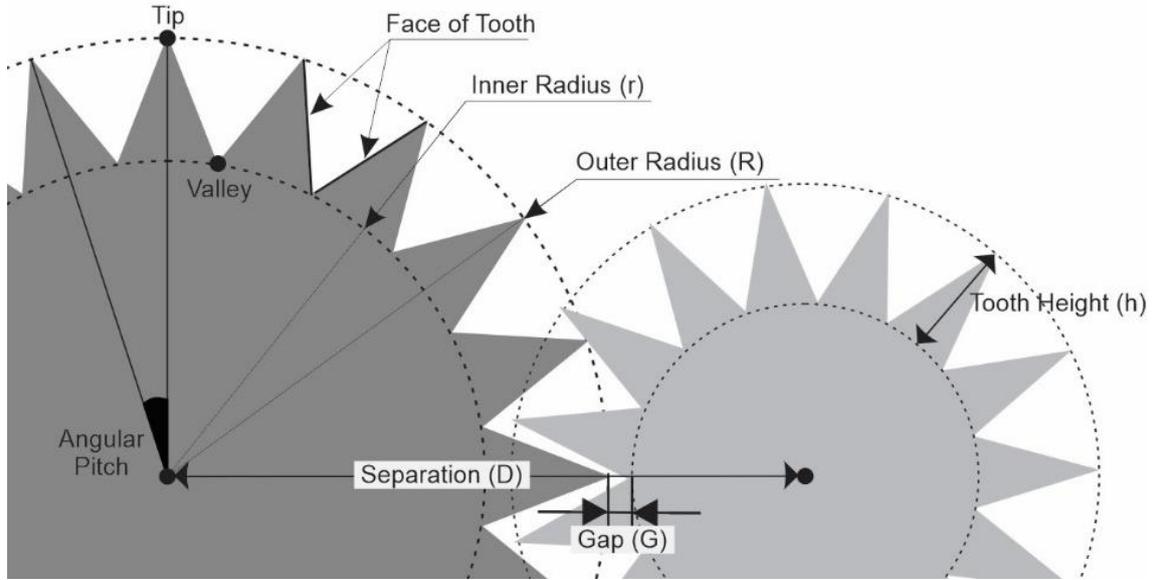

*Figure 1: Fundamental variables and geometric concepts related to triangular-tooth gears throughout the article.*

In Appendix I, it is shown how, based on these variables, a relationship can be derived to describe the angular position of the driven gear as a function of the driving gear, which we will denote as $\theta(\phi)$. This allows us to illustrate the behavior of this type of tooth, as seen in Figure 2_A, which was generated for two gears with $N1 = N2 = 8$, $R = 200$, and $r = 100$, with a separation of $D = 310$, all expressed in arbitrary length units.

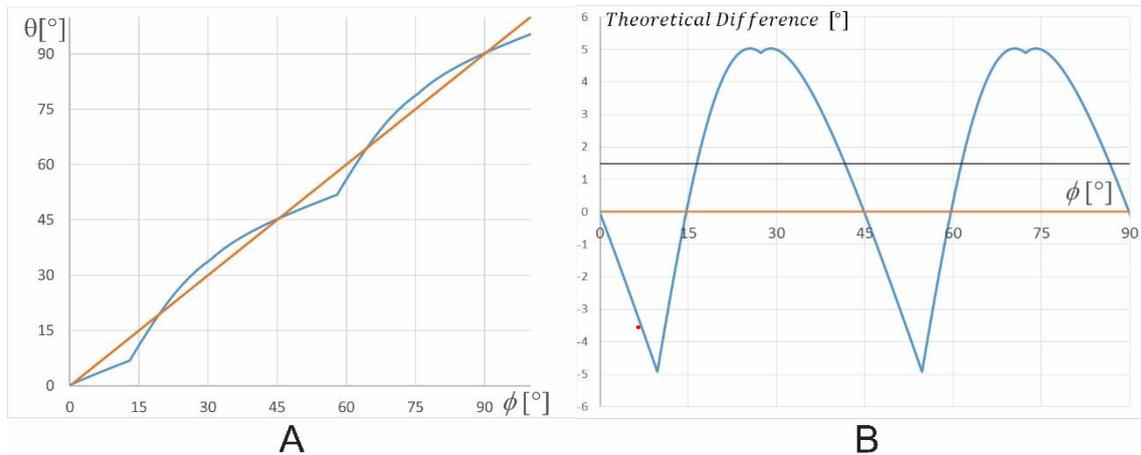

*Figure 2 A and B—In Figure A, an example of the behavior of $\theta(\phi)$ is shown for two triangular-tooth gears with a gear ratio of one. Figure B represents the Theoretical Difference, a function obtained by subtracting $\theta(\phi)$ from the ideal behavior given by Equation [1].*

From this graph, it can be observed that the triangular tooth shape causes a deviation from the expected behavior (orange line in Figure 2_A) for a pair of gears with a unitary gear ratio. One way to better appreciate this deviation is through the graph in Figure 2_B, which represents the difference between the actual position of the gear and its expected value according to Equation [1]. We refer to this as the Theoretical Difference, which is obtained by performing the following subtraction:

$$Theoretical\ Difference = \theta(\phi) - \frac{N2}{N1}\ \phi \quad [2]$$

In Figure 2_B, the black line represents the average around which a hypothetical pointer connected to the driven gear would oscillate. If we consider this average and define a symmetric range of one standard deviation on each side, the probability that a pointer observation falls within this interval will exceed 68%. Throughout this study, we will use the standard deviation

($\sigma$) as the primary criterion to assess uncertainty and quantify the deviation or displacement of the final pointers in a mechanism constructed with gears of this type.

There is no single oscillation pattern for a given pair of gears. The Theoretical Difference depends on the Initial Angles of the gears. These angles are measured from the line connecting the centers of both gears, are positive in the direction of each gear's rotation, and will be denoted as $\theta_o$ and $\phi_o$. Both Initial Angles determine the position of the function $\theta(\phi)$ relative to the origin (0,0) and, consequently, cause the Theoretical Difference to shift along both the x-axis and y-axis. In Figure 3, we can observe how the Theoretical Difference changes when starting from different Initial Angles. In Figure 3_A, it is possible for the pointer to remain almost entirely ahead at all times, or conversely, lag behind, as seen in Figure 3_C. According to our criterion, the optimal Initial Angle is the one that positions the Average at zero, as shown in Figure 3_B, ensuring that the pointer has an equal probability of being ahead or behind. The Initial Angles are values contained within the angular pitch; values greater than this repeat the contact in a cycle equal to the angular pitch.

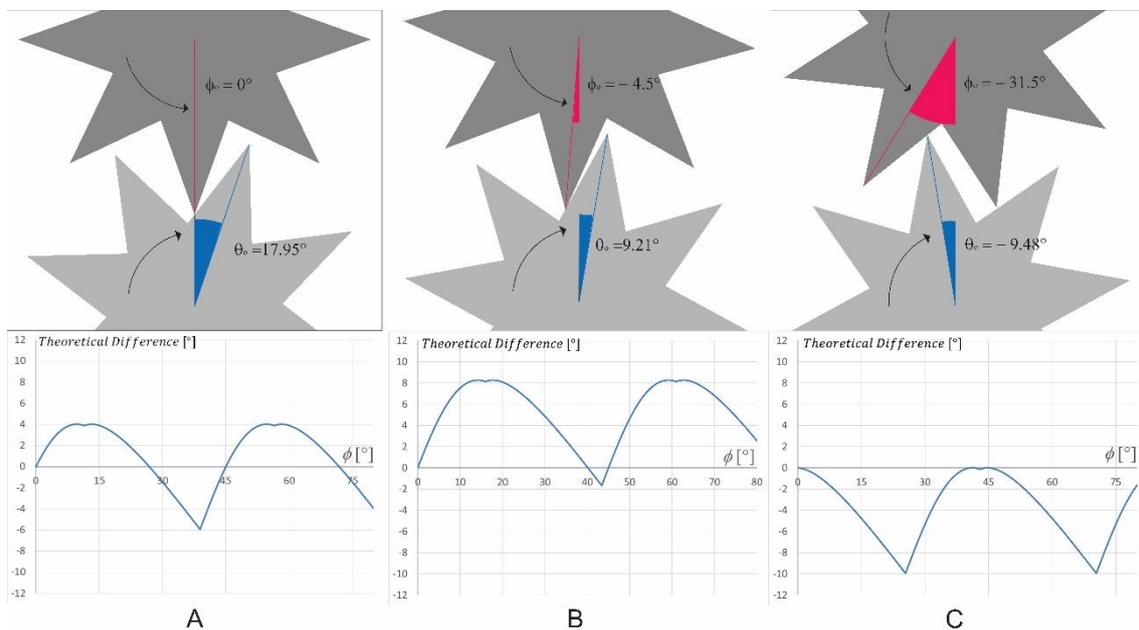

*Figure 3: In the upper graph, the Initial Angle is shown along with its value. These graphs were generated based on the same model gears discussed in this section.*

To quantify the deviation of a pointer, we use two measurements: the Standard Deviation and the Average of the oscillation. The combination of these two measurements, either by adding or subtracting them, defines the range within which the pointer is expected to deviate from the theoretical value with a probability of 68%. So far, we have assumed that this deviation from the expected values is exclusively due to the triangular tooth profile. However, later in this study, we will explore how other factors, such as manufacturing errors in the gears, can significantly influence the Theoretical Difference.

### 3- The computational model for studying a gear train

Alan Thorndike's mathematical method provides good results for two gears, but when applied to extensive gear trains like those in the Antikythera Mechanism, it requires the composition of multiple functions of the form $\theta(\phi)$. Instead of this approach, a computational simulation was developed to model and analyze gear interactions more directly. To achieve this, we created a Python program designed to simulate the contact between triangular teeth. The software is based on detecting intersections between the areas of the gears and takes the following parameters as input for each gear: the number of teeth ($N$), the outer radius ($R$), the inner radius ($r$), and the distance between each pair of gears ($D$).

The objective is to observe the Theoretical Difference of the five pointers in the Antikythera Mechanism, assuming that the gears are ideal and free from manufacturing errors. To achieve this, certain variables were adjusted in the simulation to provide a more accurate representation of the mechanism. The number of teeth ($N$) is taken from the supplementary material of the fundamental study by Freeth *et al.* (2006). The values of outer radius ($R$) are kept as close as possible to those provided by Freeth *et al.* (2006), maintaining the proportion $R_1/R_2 = N_2/N_1$. The tooth height $h$ of a gear is given by $h = R - r$. We assume that $h_1$ and $h_2$ are equal for two interconnected gears[3]. Considering an average tooth height of 1.1 mm for the entire Antikythera Mechanism, the inner radius $r$ all gears is defined based on $R$.

If the center distance $D$ is set to $R1 + R2 - h$, the tip of one gear would touch the valley of the other, causing a jam. Therefore, $D$ must be greater than $R1 + R2 - h$, specifically $R1 + R2 - h + G$, where the clearance $G$ was previously introduced in Figure 1. Typically, $G$ is expressed as a percentage of $h$. When $G$ is less than 10% of $h$, jamming occurs, and when it exceeds 90% of $h$, disengagement happens. Acceptable values for $G$ are around 40% (see Appendix II), which is close to the optimal value for preventing both disengagement and jamming, even in the presence of manufacturing errors in the gears.

In the computational model developed in Python, a random variation of the Initial Angle was also included, as we assume that this parameter could not have been consciously controlled in the Antikythera Mechanism. In other words, we propose that the maker was unable to determine in advance how to position the gears so that their teeth would initially align with specific Initial Angles, and no adjustment mechanism for this variable has been identified in the device. For example, if all gear pairs contributed to a delay effect in the deviation, as observed in Figure 3_B, the final pointer of the gear train would return values lower than those expected according to the overall transmission ratio. Therefore, it is more likely that the initial contact was assigned randomly, which is why we use a uniform distribution within the angular pitch range of each driving gear.

In the simulation, the pin-slot device was deliberately excluded because it does not influence the phenomenon under analysis—the impact of the triangular tooth profile. In other words, its presence would not provide relevant information for the study, so omitting it allows the focus to remain on the factors that genuinely affect the system's behavior. This approach helps simplify the simulation and yields clearer and more precise results.

Figure 4 illustrates one of the thirty instances of the three main pointers of the Antikythera Mechanism: the Lunar Pointer (Figure 4_A), the Metonic Pointer (Figure 4_B), and the Saros Pointer (Figure 4_C), plotting the corresponding Theoretical Difference for each in degrees as a function of $\phi$ in degrees during the first revolution of the Solar Pointer.

---

[3] In a study conducted by Efstathiou et al. (2012) on the measurement of the geometric parameters of the Antikythera Mechanism's gears, it was observed that the average tooth height is 1.1 mm. This value will be used as a reference in this study for all gear tooth heights.

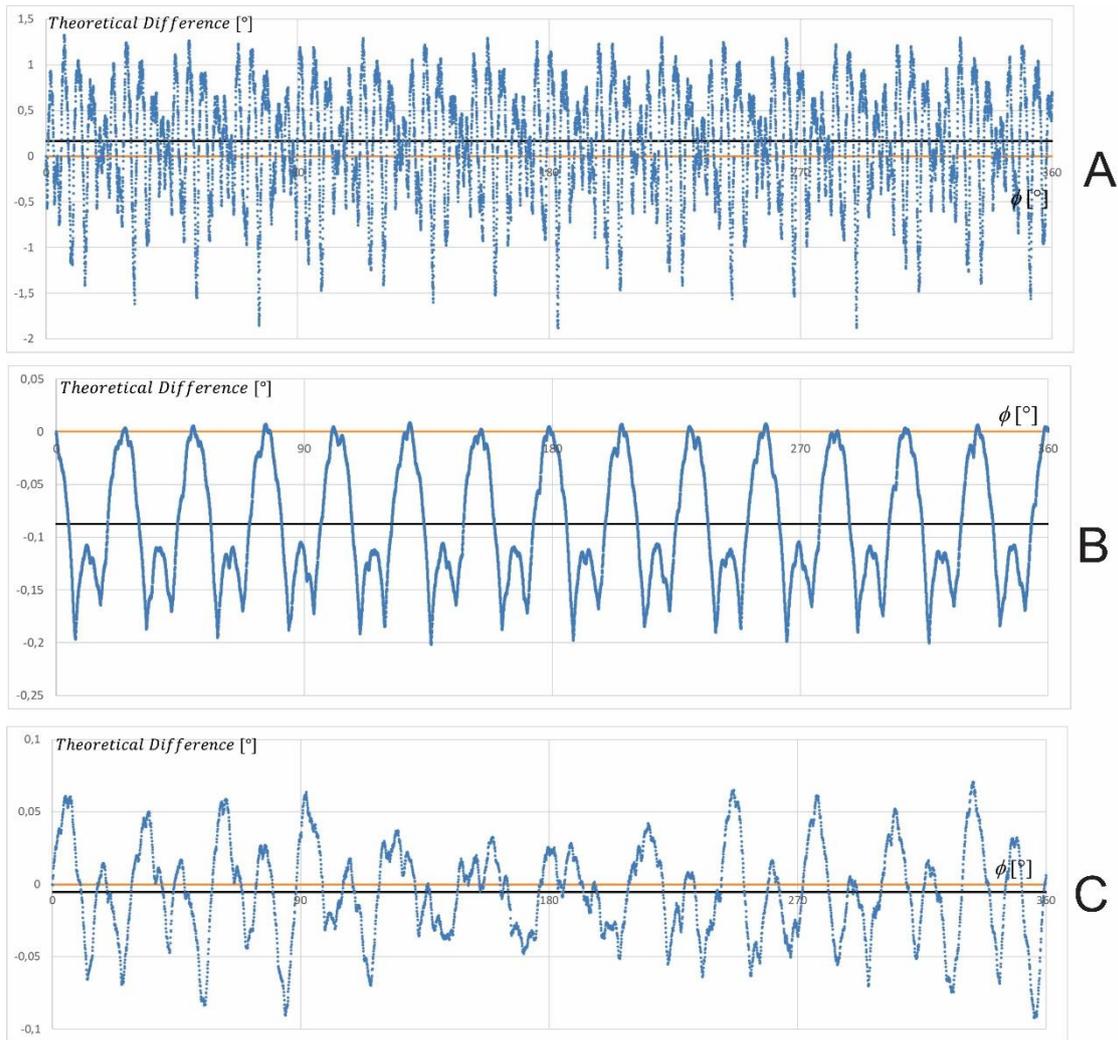

*Figure 4. Behavior of the Lunar Pointer (A), Metonic Pointer (B), and Saros Pointer (C) within one revolution of the Solar Pointer.*

Table 1 presents the results. For each gear train of the indicators, the standard deviation ($\sigma$), the average maximum deviation in each trial ($Dm$), and the average absolute value of the oscillation mean ($P$) are provided after all gears in each train have completed at least one full rotation.

|  | Lunar Pointer [°] | Metonic Pointer [°](days) | Eclipse Pointer [°](days) | Game Pointer [°] | Exeligmos Pointer [°] |
|---|---|---|---|---|---|
| $\sigma$ | 0.60 | 0.053(0.20) | 0.038(0.087) | 0.059 | 0.051 |
| $Dm$ | 1.84 | 0.14(0.54) | 0.14(0.32) | 0.18 | 0.13 |
| $P$ | 0.66 | 0.044(0,17) | 0.029(0.066) | 0.038 | 0.042 |

*Table 1. Standard deviation ($\sigma$), maximum deviation ($Dm$ ), and average ($P$) of the different gear trains for triangular-tooth gears without manufacturing errors. In the case of the Lunar Pointer, the program was executed until gear B2 completed a full rotation. For the other pointers, the following criteria were set: the Metonic Pointer completed 3.8 revolutions of the Solar Pointer, while the Saros Pointer completed 4.5 revolutions of the Solar Pointer. This ensures that all gears in each train have rotated at least once.*

The 30 graphs obtained from the computational modeling of each gear train differ from the examples shown in Figures 4_A, 4_B, and 4_C. This difference arises because the Initial Angles in each gear pair were assigned randomly, resulting in unique variations in each graph. For this reason, Table 1 presents the results as averages of the 30 trials conducted after a full operating cycle for each pointer.

Based on the values in Table 1, we can conclude that the Antikythera Mechanism would have introduced differences from the Transmission Ratio results of each pointer that are fairly acceptable. For example, if the Lunar anomaly introduced by the pin-slot device has an angular variation of $\pm6.53°$ (Freeth *et al.*, 2006), then a standard deviation of $\pm0.60°$ would have allowed for some harmonic variation to be observed in the Lunar Pointer. However, this variation could deviate from the true theoretical value of the Moon by up to nearly 2.5°, given that $1.84° + 0.66° = 2.5°$. This indicates that 68% of the observations would still have been acceptable, with deviations smaller than the maximum variation introduced by the pin-slot device.

In the Saros and Metonic dials, Table 1 includes both the values in degrees and their equivalents in days. If these pointers were associated with a spiral dial containing compartments representing each lunar month of the cycle, then the pointer's accuracy would be expected to fall within the duration of a lunar month. The results of this ideal model show that the deviation of these pointers was precise, never deviating from the expected value by more than a day: $0.74$ days ($0.20\ d\ +\ 0.54\ d$) in the Metonic train and approximately $0.41$ days ($0.32\ d\ +\ 0.087\ d$) in the Saros train.

## 4- <u>Introducing manufacturing errors in the gears</u>

In the previous section, we assumed that the gears with triangular teeth were perfectly constructed. However, this is not the case for the gears of the Antikythera Mechanism. Edmunds identified two sources of manufacturing errors: randomness in the distribution of the teeth in all the gears and a certain systematic error in some of them.

He expresses the random error in the distribution of the teeth as a fraction, representing the standard deviation of the angular difference between the ideal and actual positions of the tip of each gear tooth, divided by the angular pitch of the teeth in degrees. For example, if he found a standard deviation error of 1° in a gear with 60 teeth (with an angular pitch of 6°), the fractional error is 1°/6° = 0.16. This type of error was found in all the gears of the Antikythera Mechanism. Edmunds established a variation range for his study, which ranged between 0.04 and 0.08.

Additionally, he detected a systematic error in some gears, consisting of a sinusoidal shift in the position of the teeth. This error could be explained by two possible causes. One possible cause is the eccentricity of the actual center of the gear (i.e., the center around which the gear rotates) relative to the geometric center (i.e., the center of the circle that best fits the tips or valleys of all the teeth). However, in his 2011 study, Edmunds could not conclude that this was exclusively due to this phenomenon. He also attributed it to the shift generated when marking the tooth positions on a bronze disc. The systematic error is expressed by the amplitude in degrees of the sinusoidal error function.

Based on these errors, Edmunds developed four possible scenarios (from Case A to Case D), combining plausible values for these errors, with the tooth error ranging between 0.04 and 0.08, and the systematic error between 0° and 2°. He applied his model to the gear trains of the Lunar, Metonic, and Saros indicators. The description of the input errors for each case and the results obtained are summarized in Table 2.

| | Input | | Output | | |
|---|---|---|---|---|---|
| | Eccentricity error [°] | Tooth error | Lunar Pointer [°] σ - Dm | Metonic Pointer [°](days) σ - Dm | Eclipse Pointer [°](days) σ - Dm |
| CASE A | 0 | 0,04 | 4,4 -13 | 0,32(1,2d) -1,15 (4,4d) | 0,38 (1,7d)-1,3(6d) |
| CASE B | 0,50 | 0,04 | 7-20 | 0,5 (2d) - 1,5 (6 d) | 1,1 (5d) - 2,7 (12d) |
| CASE C | 1,00 | 0,04 | 14 – 33 | 0,85 (3d) - 2,3 (9 d) | 2,1 (10d)- 4,9 (22d) |
| CASE D | 2,00 | 0,08 | 26 – 57 | 1,6 (6d) - 3,9 (15d) | 3,8 (17d) - 8,6 (39d) |

*Table 2: Summary of input errors and output deviations for Edmunds' model*

The adaptation of these manufacturing errors to our initial model of perfect teeth was carried out as follows: the position of the tooth tips was assigned the normalized random variation proposed by Edmunds. The valleys were positioned symmetrically between the two tip positions, already affected by the error. We chose to divide the systematic error into two parts: 50% of the error affects the sinusoidal distribution of the tip positions (with the valleys positioned in the same way, symmetrically between two positions already affected by the error), and the remaining 50% affects the eccentricity of the gear axis.

This last 50% of the systematic error can be incorporated into our model concretely by rotating the gear around a center other than the geometric one. This causes the gear to behave like a cam, oscillating and modifying the G parameter between gears—something that was previously constant in our ideal tooth model.

The remaining 50% of the systematic error affects the distribution of the tooth positions. For example, a 5° error in our model is divided into 2.5° of amplitude in the sinusoidal distribution of the teeth, while the other 2.5° results in an axis shift of approximately 4 millimeters in a gear with a 10-centimeter radius.

As we mentioned, Edmunds' analysis does not take into account the triangular shape of the gear teeth. Therefore, we introduced both random and systematic errors into our model and found values that do not significantly differ from those of Edmunds. The deviation in the model that includes triangular teeth is sometimes slightly greater and sometimes slightly smaller, oscillating around Edmunds' error.

Figure 5_A illustrates the deviation produced by both models for a pair of 50-tooth gears (R=200, r=180, and G=50%) with a random error of 0.05 and a systematic error of 0.6°, concentrated exclusively on the axis shift. The model was also tested for different percentages of the systematic error, showing a similar behavior in all cases. In Figure 4_B, we see the same gear pair with the same random and systematic errors, but this time with 50% affecting the distribution and the other 50% affecting the axis shift.

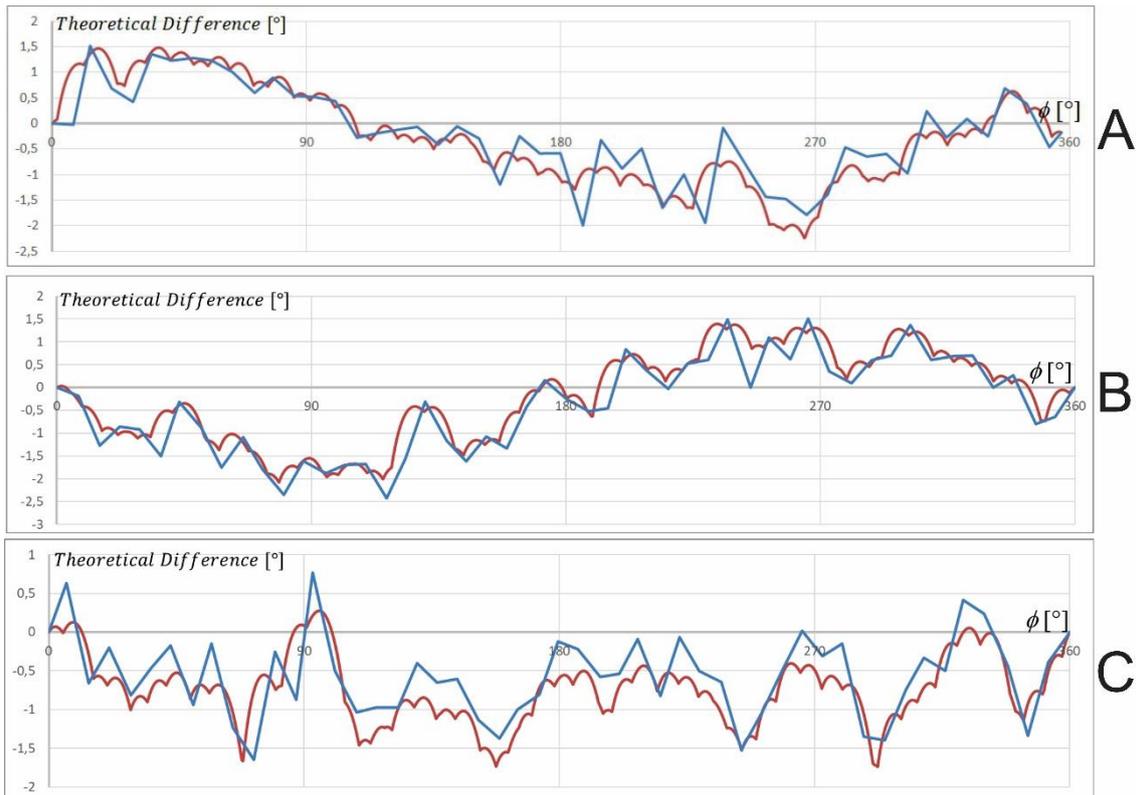

*Figure 5: _A: Random error of 0.05 and systematic error of 0.6°, with the entire error applied to the eccentricity and none to the sinusoidal distribution of the angular positions of the teeth. _B: Random error of 0.05 and systematic error of 0.6°, applied 50-50% to each cause. C: Random error of 0.05 and systematic error of 0.6°, fully applied to the sinusoidal distribution of the tooth positions in each gear.*

What these three graphs in Figure 5 demonstrate is that changing the percentage of systematic error applied to one cause or the other does not produce a significant effect on the final behavior of a hypothetical pointer. This leads us to conclude that using the 50-50% distribution in our considerations is a valid hypothesis, although not necessarily an accurate representation of how the gears of the Antikythera Mechanism were actually manufactured. Similarly, Edmunds was unable to confirm this; he could only determine that some gears exhibited a systematic error between 0.5° and 1°. Based on this, he decided to adopt this criterion for the development of each case, which led him to conclude that the Antikythera Mechanism could have functioned better than Case C and probably as well as Case B.

### 5- The consequences for the Antikythera Mechanism of studying the triangular tooth profile

When we introduced the errors defined by Edmunds' Cases A to D into our Python computational model, the uneven distribution of the teeth and the change in the G spacing caused by the eccentricity—accounting for 50% of the systematic error—led to several instances of jamming and disengagement in the interconnected gears. To quantify these effects, we analyzed each interconnected gear pair separately in the four scenarios (Cases A to D) with a G spacing of 40%, running the program 200 times for each pair in each scenario and recording occurrences of jamming and disengagement. Our program was designed to complete one full rotation of the gear with the most teeth, allowing the driven gear to rotate freely if a jam or disengagement was detected until movement was possible again. Once movement resumed, the pair continued operating, searching for further instances of jamming or disengagement. Table 3 presents the results for each gear pair, where the first number represents the number of trials with at least one jam, and the second indicates the trials where at least one disengagement occurred. Cells with one or fewer trials showing jamming or disengagement are shaded.

| Pointer Pairs / Case | Luna | | | | | | Metonic and Saros | | | | | | Games | Exeligmos | |
|---|---|---|---|---|---|---|---|---|---|---|---|---|---|---|---|
| | b2-c1 | c2-d1 | d2-e2 | e5-k1 | k2-e6 | e1-b3 | b2-l1 | l2-m1 | m3-e3 | e4-f1 | f2-g1 | m2-n1 | n2-o1 | g2-h1 | h2-i1 |
| A | 0/0 | 0/0 | 0/0 | 0/0 | 0/0 | 0/0 | 0/0 | 0/0 | 2/0 | 0/0 | 0/0 | 2/0 | 0/0 | 0/0 | 2/0 |
| B | 0/0 | 0/0 | 0/0 | 0/0 | 0/0 | 0/0 | 0/0 | 1/0 | 168/0 | 99/0 | 0/0 | 5/0 | 0/0 | 1/0 | 8/0 |
| C | 5/0 | 1/0 | 197/0 | 5/0 | 10/0 | 1/0 | 6/0 | 118/0 | 200/1 | 200/0 | 7/0 | 10/0 | 10/0 | 4/0 | 17/0 |
| D | 200/0 | 195/0 | 151/49 | 185/0 | 179/0 | 148/0 | 200/0 | 68/172 | 0/200 | 105/95 | 198/0 | 200/0 | 176/0 | 200/0 | 200/0 |

*Table 3: Jamming and disengagement results for the four scenarios defined by Edmunds. A total of 200 trials were conducted for each gear, distributing the systematic error with 50% affecting the tooth distribution and the other 50% affecting the axis shift. Shaded cells in the table indicate cases where the jamming percentage was below 1%. The notation for the Antikythera Mechanism gear pairs used throughout this study follows Freeth et al. (2006). The shaded cells correspond to situations where the 1% blockage criterion has not been exceeded.*

It is important to note that a single jam would stop the entire mechanism, while a single disengagement would partially halt the dependent indicators, introducing critical synchronization errors. We adopted a failure rate threshold of 1%[4] for each gear pair (meaning that in 200 trials, a maximum of 2 jams is acceptable), which corresponds to the shaded cells in Table 3. The table clearly demonstrates that Cases C and D result in such a high number of jamming and/or disengagement events that, if the mechanism had these manufacturing errors, it would become unusable. Case A presents no issues, while Case B is also likely too problematic due to two specific gear pairs (M3-E3 and E4-F1). The tolerable error range would probably fall between the conditions of Case A and Case B.

Table 3 also shows that gear pairs in which one of the gears has a large number of teeth and, therefore, a large radius (such as D2-E2, M3-E3, E4-F1, and L2-M1) tend to produce more jamming and/or disengagement events. This occurs because 50% of the systematic error affects the eccentricity, which is proportional to the radius, while the tooth height remains constant. As a result, the G spacing varies significantly with the same eccentricity error in larger gears, increasing the likelihood of jamming and disengagement. Assigning the same eccentricity error, as defined by Edmunds, to all gears regardless of their size results in a greater distance between the geometric and actual centers in larger gears. Therefore, it is reasonable to independently analyze the maximum tolerable error for each gear pair.

In this section, we analyze the maximum tolerable random and systematic errors (still considering it as 50% affecting the tooth distribution and 50% affecting the eccentricity) for each gear pair. First, we assume there is no systematic error and gradually increase the error in the teeth.

---

[4] This criterion has been adopted by the authors of this article based on the following reasoning: if we assume a jamming rate of 1% for a single gear pair and consider that the Antikythera Mechanism consists of 15 gear pairs, then, assuming failures occur independently for each pair, the probability of the entire mechanism failing would be approximately 14%. In contrast, if each gear pair had a 5% failure probability, the total probability of malfunction would rise to nearly 54%. This would mean that out of every ten mechanisms produced, approximately half would be unable to move their indicators—an outcome clearly unacceptable for any manufacturer.

| Pointer Pairs | Lunar | | | | | | Metonic and Saros | | | | | | Games | Exeligmos | |
|---|---|---|---|---|---|---|---|---|---|---|---|---|---|---|---|
| Errors | b2-c1 | c2-d1 | d2-e2 | e5-k1 | k2-e6 | e1-b3 | b2-l1 | l2-m1 | m3-e3 | e4-f1 | f2-g1 | m2-n1 | n2-o1 | g2-h1 | h2-i1 |
| 0.035 | 0/0 | 0/0 | 0/0 | 0/0 | 0/0 | 0/0 | 0/0 | 0/0 | 0/0 | 0/0 | 0/0 | 2/0 | 0/0 | 0/0 | 0/0 |
| 0.040 | 0/0 | 0/0 | 0/0 | 0/0 | 0/0 | 0/0 | 0/0 | 0/0 | 1/0 | 0/0 | 0/0 | 0/0 | 0/0 | 0/0 | 2/0 |
| 0.045 | 0/0 | 1/0 | 1/0 | 0/0 | 0/0 | 0/0 | 0/0 | 0/0 | 10/0 | 1/0 | 0/0 | 11/0 | 0/0 | 2/0 | 10/0 |
| 0.050 | 1/0 | 4/0 | 10/0 | 2/0 | 0/0 | 4/0 | 1/0 | 2/0 | 17/0 | 5/0 | 3/0 | 20/0 | 1/0 | 5/0 | 20/0 |
| 0.055 | 1/0 | 8/0 | 14/0 | 5/0 | 4/0 | 10/0 | 2/0 | 8/0 | 36/0 | 13/0 | 6/0 | 40/0 | 0/0 | 12/0 | 46/0 |
| 0.060 | 17/0 | 16/0 | 28/0 | 12/0 | 15/0 | 15/0 | 10/0 | 15/0 | 68/0 | 18/0 | 22/0 | 60/0 | 4/0 | 33/0 | 66/0 |
| 0.065 | 20/0 | 20/0 | 49/0 | 21/0 | 24/0 | 19/0 | 24/0 | 30/0 | 94/0 | 52/0 | 28/0 | 87/0 | 17/0 | 53/0 | 95/0 |

*Table 4: Jamming / decoupling results for varying tooth errors in the absence of systematic error. This was also done for 200 trials. The shaded cells correspond to situations where the 1% blockage criterion has not been exceeded.*

Table 4 presents the results of varying the tooth error from 0.04 to 0.65 in increments of 0.005 while keeping the systematic error at zero. We conducted 200 trials for each gear pair. In this scenario, only jamming events were observed, which is expected given the absence of systematic error. Notably, the tolerance to tooth errors varied among different gear pairs. For example, the M2-N1 pair could not tolerate errors greater than 0.04, while the B2-C1 pair continued to function without issues even with errors up to 0.055. If a uniform maximum tooth error is required for all gear pairs, it should not exceed 0.04.

Next, while maintaining the maximum tolerable random error for each pair, we increased the eccentricity error. Table 5 shows the results of varying the systematic error (distributed 50%-50%) from 0.1° to 1.1° in increments of 0.1°. For each pair, we conducted 200 trials. As before, the first number in each cell indicates the number of trials with at least one jamming event, while the second number represents the trials with at least one disengagement event. While some gear pairs, such as E1-B3, can tolerate a systematic error of up to 0.6°, others, like D2-E2, can only function with errors up to 0.1°. If a uniform maximum eccentricity error is required for all gear pairs, it should not exceed 0.1°.

| Pointer Pairs | Lunar | | | | | | Metonic and Saros | | | | | | Games | Exeligmos | |
|---|---|---|---|---|---|---|---|---|---|---|---|---|---|---|---|
| E. Max. E. Sis. [°] | b2-c1 | c2-d1 | d2-e2 | e5-k1 | k2-e6 | e1-b3 | b2-l1 | l2-m1 | m3-e3 | e4-f1 | f2-g1 | m2-n1 | n2-o1 | g2-h1 | h2-i1 |
| | 0.055 | 0.045 | 0.045 | 0.050 | 0.050 | 0.045 | 0.055 | 0.05 | 0.040 | 0.045 | 0.045 | 0.040 | 0.055 | 0.045 | 0.040 |
| 0.1 | 1/0 | 1/0 | 0/0 | 1/0 | 0/0 | 1/0 | 1/0 | 0/0 | 0/0 | 1/0 | 0/0 | 1/0 | 0/0 | 0/0 | 1/0 |
| 0.2 | 0/0 | 0/0 | 3/0 | 1/0 | 2/0 | 0/0 | 1/0 | 1/0 | 2/0 | 1/0 | 1/0 | 1/0 | 1/0 | 1/0 | 1/0 |
| 0.3 | 0/0 | 2/0 | 4/0 | 3/0 | 3/0 | 0/0 | 0/0 | 2/0 | 12/0 | 10/0 | 2/0 | 0/0 | 1/0 | 0/0 | 1/0 |
| 0.4 | 1/0 | 1/0 | 6/0 | 3/0 | 5/0 | 0/0 | 1/0 | 1/0 | 41/0 | 45/0 | 2/0 | 5/0 | 0/0 | 1/0 | 2/0 |
| 0.5 | 2/0 | 1/0 | 14/0 | 9/0 | 7/0 | 2/0 | 2/0 | 1/0 | 163/0 | 146/0 | 0/0 | 3/0 | 1/0 | 2/0 | 0/0 |
| 0.6 | 2/0 | 3/0 | 33/0 | 7/0 | 8/0 | 1/0 | 1/0 | 11/0 | 200/0 | 196/0 | 2/0 | 3/0 | 1/0 | 1/0 | 1/0 |
| 0.7 | 1/0 | 5/0 | 80/0 | 8/0 | 10/0 | 3/0 | 1/0 | 25/0 | 200/0 | 200/0 | 1/0 | 5/0 | 5/0 | 1/0 | 1/0 |
| 0.8 | 13/0 | 5/0 | 141/0 | 16/0 | 15/0 | 4/0 | 6/0 | 62/0 | 200/0 | 200/0 | 1/0 | 4/0 | 7/0 | 3/0 | 1/0 |
| 0.9 | 11/0 | 4/0 | 186/0 | 13/0 | 11/0 | 5/0 | 9/0 | 107/0 | 200/0 | 200/0 | 9/0 | 6/0 | 11/0 | 5/0 | 4/0 |
| 1.0 | 16/0 | 11/0 | 198/1 | 23/0 | 25/0 | 4/0 | 16/0 | 147/0 | 200/0 | 200/3 | 11/0 | 7/0 | 22/0 | 9/0 | 4/0 |
| 1.1 | 31/0 | 9/0 | 193/7 | 33/0 | 31/0 | 5/0 | 24/0 | 166/0 | 200/189 | 165/35 | 17/0 | 14/0 | 29/0 | 6/0 | 7/0 |

*Table 5: Jamming/decoupling results varying the eccentricity errors (E. Sis [°]) and keeping the maximum tolerable random tooth error as indicated in row 3 (E. max). As in Tables 2 and 3, the shaded cells correspond to situations where the 1% blockage criterion has not been exceeded.*

Edmunds measured the random tooth error for each existing gear in the computed tomography scans of the mechanism. Therefore, we can compare our maximum tolerable random error with the actual errors found in the mechanism. Since our unit of analysis is the gear pair, we assigned the measured errors to both gears in each pair.

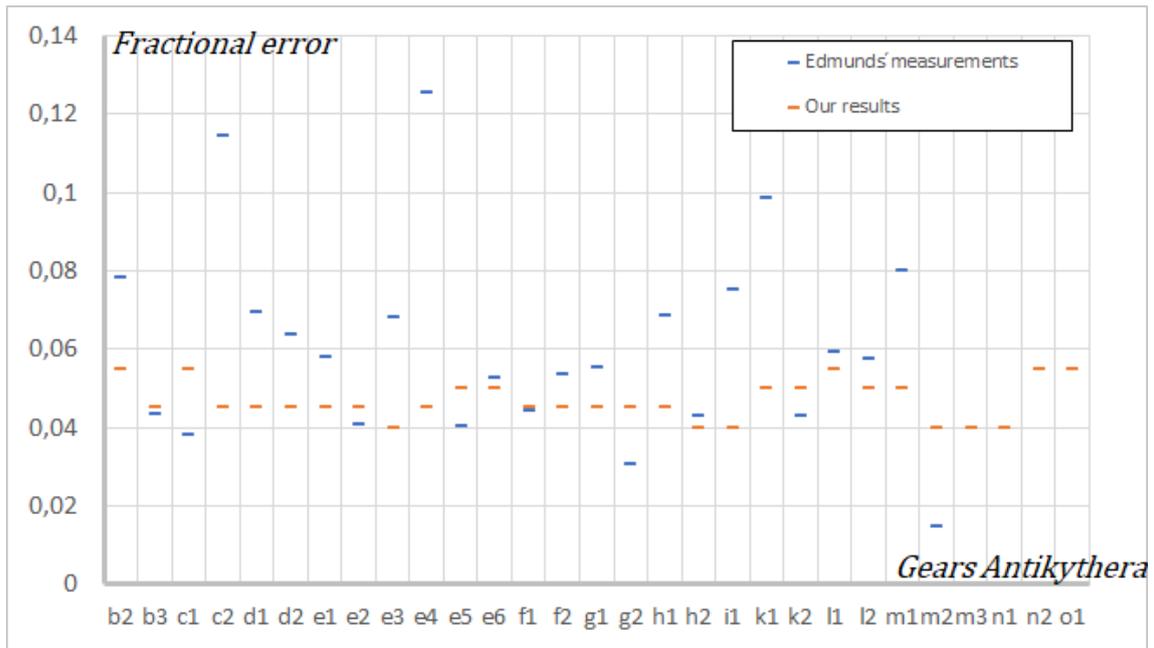

*Figure 6: Comparative graph between the pitch errors measured by Edmunds in his 2011 study and the maximum tolerable errors obtained in our computational model. The gears m3, n1, n2, and o1 do not have values measured by Edmunds because they were not found in the fragments.*

Figure 6 illustrates the comparison between our maximum tolerable random error, estimated based on the jamming limits that restrict the operation of the MA, and the errors measured by Edmunds in his 2011 study. In this case, the actual gear errors typically exceed the tolerable limits. This indicates that if the Antikythera Mechanism had the errors identified by Edmunds, it would not function properly. We simulated a model 30 times using Edmunds' error values and an optimal separation (G) of 40%, which resulted in jamming in 90% of instances. On average, the Lunar train jammed after 129 days, the Metonic and Games train operated for 234 days before stopping, and the Saros and Exeligmos train functioned without jamming for 121 days. While no train fully decoupled under Edmunds' values, they were unable to rotate freely to perform their intended predictions.

On the other hand, as we have demonstrated, a functional model would tolerate errors between those of Case A and Case B, likely closer to Case A. Since the triangular shape of the teeth does not introduce significant errors, if the Antikythera Mechanism were a functional device, its indicators would produce errors within the range found by Edmunds for Cases A and B. We then set out to construct a mechanism using the tolerable errors we identified, resulting in an Antikythera Mechanism that exhibits the behavior shown in Table 6. As we can see, it still functions very similarly to Case B, as confirmed by Edmunds in his 2011 study. This reinforces the idea that the Antikythera Mechanism, despite its manufacturing imperfections, could have operated within an acceptable margin of error, aligning with previous findings.

| | Lunar Pointer [°] | Metonic Pointer [°](days) | Eclipse Pointer [°](days) | Game Pointer [°] | Exeligmos Pointer [°] |
|---|---|---|---|---|---|
| σ | 9.51° | 0.37° | 0.46° | 0.47° | 0.14° |
| Dm | 23.12° | 0.90° | 1.09° | 1.17° | 0.30° |
| P | 12.22° | 0.40° | 0.56° | 0.64° | 0.22° |

*Table 6: Standard deviation (σ), maximum deviation (Dm), and average (P) of the different gear trains for gears with triangular teeth, considering the maximum admissible random and systematic errors (50%-50%). Averages were calculated over 30 trials, and the cycles considered ensured that all gears completed at least one full rotation.*

## 6- **Conclusiones**

We developed a computational model that allowed us to incorporate the effect of the triangular tooth shape, examined by Thorndike, into the manufacturing errors analyzed by Edmunds. First, we found that the triangular shape in ideal teeth introduces deviations that are almost negligible compared to those identified by Edmunds, being on average seven times smaller. Second, when we introduced tooth errors and eccentricity into our model of gears with triangular teeth, the theoretical deviations of the indicators did not differ significantly from those found by Edmunds. However, the occurrence of jamming and disengagement events limits the size of tolerable errors: excessive eccentricity or significant errors in tooth distribution would either cause the mechanism to stop completely due to jamming or introduce critical desynchronization among the indicators, potentially leading to the total failure of some indicators due to disengagement. We studied the maximum error tolerance for each gear pair and compared it with the values Edmunds found in the computed tomography scans. In most cases, the errors identified in the mechanism exceed the tolerable limits. A mechanism with the errors found by Edmunds would generally stop after the solar pointer completed one-third (120°) of its full rotation. Other manufacturing errors not considered in this analysis could further complicate the situation. For example, we assumed (1) that the gears were perfectly circular (i.e., the distance from the geometric center to the tip and valley of each gear was uniform), (2) that the sides of the triangular teeth were straight lines, (3) that the V-shape of each tooth was symmetrical, (4) that the gears were perfectly flat and each pair operated in the same plane, (5) that the ratio of gear sizes matched exactly with the ratio of tooth counts, and so on.

As a result, two possibilities arise: either the mechanism never functioned, or its errors were smaller than those found by Edmunds. While it is conceivable—though unlikely—that someone would go to the trouble of constructing such a complex yet non-functional device, there are strong reasons to question whether Edmunds' values accurately represent the mechanism's original errors. The effects of 2,000 years underwater likely caused corrosion that may have deformed the gears, while the resolution of the computed tomography scans might not be sufficient to precisely detect the tips or valleys of the teeth. Additionally, since many gears are partially destroyed, Edmunds had to work with incomplete data. For these reasons, it is reasonable to believe that the errors identified by Edmunds may be exaggerated compared to the mechanism's original state.

This analysis suggests that we must be cautious in assuming that our measurements of the fragments perfectly reflect their original values. Instead, it highlights the need for further research and the possible development of more refined techniques to better understand the true accuracy and functionality of the Antikythera Mechanism.

We conclude that the Antikythera Mechanism could have functioned better than Case B and slightly worse than Case A. In fact, the sub-case we defined as the functional limit exhibits errors in its final indicators that would still have made it a valid device for solving the proportionality calculations it was designed for.

## 7- **Acknowledgements**

The authors would like to thank the National University of Mar del Plata for its contributions that made this research possible. We also gratefully acknowledge the invaluable collaboration of Dr. Christián Carman, whose comments and corrections enabled this article to come to fruition.

## 8- **Bibliography**

### 9- Appendix I: Proof of the transmission ratio for gears with triangular teeth

We can identify two distinct moments when the triangular teeth come into contact: one where the driving face touches the tip of the driven gear and, conversely, when the tip of the driving gear touches the face of the driven gear. Phase 1 will be referred to as the first stage, and Phase 2 as the second stage.

Alan Thorndike, in his study on transmission through triangular teeth, focuses his analysis exclusively on Phase 2. To easily verify the expression proposed by Thorndike, we present a proof based on the diagram in Figure I.

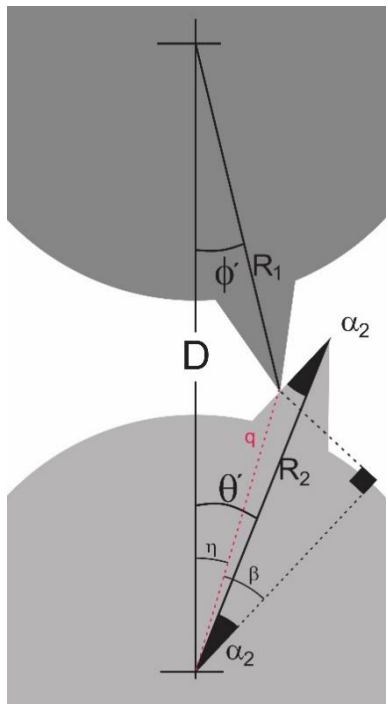

*Figure I: Moment of contact between triangular teeth where the tip of the driving gear (dark gray) touches the face of the driven gear (light gray). Here, the driving gear is rotating counterclockwise. This is referred to as Phase 2.*

The length $q$ is defined using the law of cosines, obtained in the same manner as Thorndike:

$$q = \sqrt{R_1{}^2 + Y^2 - 2R_1 Y \cos \phi'}\ .$$

The angles $\eta$ and $\phi'$ can then be related using the law of sines:

$$\frac{\sin \eta}{R_1} = \frac{\sin \phi'}{q}$$

On the other hand:

$$\sin \beta = \frac{R_2 \cdot \sin \alpha_2}{q}$$

Then, given the equality $\theta' + \alpha_2 = \eta + \beta$, the following relationship can be derived:

$$\theta'(\phi') = asin\left[\frac{R_1 \sin(\phi')}{q}\right] + asin\left[\frac{R_2 \sin(\alpha_2)}{q}\right] - \alpha_2 \qquad [I]$$

Expression [I] is valid if we consider that the position of both the driving and driven gears is measured from an angular reference system with its origin on the line connecting the gear axes, with the positive direction counterclockwise for the driving gear and clockwise for the driven gear. Alternatively, we can say that the positive direction follows the rotation of the gears as shown in Figure I.

The relationship $\theta'(\phi')$ for Phase 1 of the contact between triangular teeth requires a proof very similar to the previous one, as shown in Figure II.

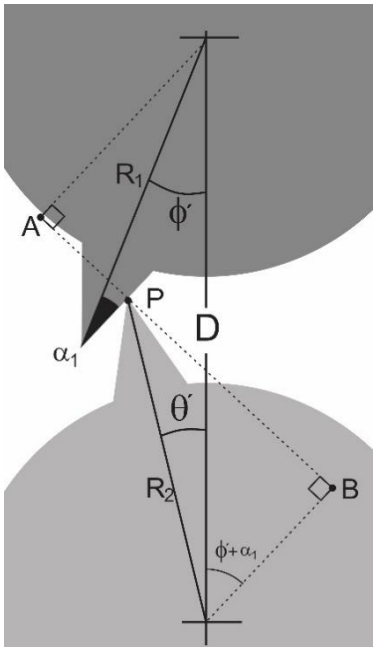

*Figure II: Moment when the driving face (dark gray gear) touches the tip of the tooth of the driven gear (light gray). In this case, the driving gear is rotating counterclockwise, corresponding to Phase 1.*

Si tomamos el segmento normal a la cara del diente del engranaje conductor, APB, podemos calcular su longitud:
If we take the segment normal to the face of the driving gear tooth, $\overline{APB}$, we can calculate its length:

$$\overline{APB} = Y \sin(\phi' + \alpha_1)$$

The lengths of segments $\overline{AP}$ and $\overline{PB}$ are:

$$\overline{AP} = R_1 \sin(\alpha_1) \quad \text{y} \quad \overline{PB} = R_2 \sin(\alpha_1 + \theta' + \phi')$$

After summing the segments $\overline{APB} = \overline{AP} + \overline{PB}$, the desired expression can be obtained. However, to ensure that it aligns with the angular reference system used in the previous expression, the signs must be adjusted, resulting in the final form:

$$\theta'(\phi') = asin\left[\frac{Y \sin(\phi' - \alpha_1)}{R_2} + \frac{R_1 \sin(\alpha_1)}{R_2}\right] - \phi' + \alpha \qquad [II]$$

To obtain the expression $\theta(\phi_o)$ mentioned at the beginning of Section II of the main text, expressions [I] and [II] must be adjusted to match the reference system adopted in the article. This requires applying the following transformations:

$$\phi = \phi' - \phi_o' \quad y \quad \theta = \theta' - \theta_o'$$

Where $\phi_o'$ and $\theta_o'$ are the initial angles measured from the line connecting the gear axes, and $\theta_o'$ is obtained as $\theta'(\phi_o')$. With these adjustments, we are now ready to determine the theoretical difference in the expression $\theta(\phi_o)$ mentioned earlier in the main text. In Figure III_A, we see the result of applying this function to the case of two identical gears with N=8 and the same dimensions as the model example used in the main article, with initial angles $\phi_o' = 0°$ and $\theta_o' = \theta'(0°) = 17.96°$. Figure III_A shows an intersection between the two functions, representing the moment of transition from Phase 1 to Phase 2. To assign the correct width to this function and predict when subsequent transitions occur, a numerical approach is used to find the point where an angular step width produces a matching ordinate value. As seen in **Figure III_B**, this trimmed segment can then be extended, as it correctly follows the complete cycle of the theoretical difference.

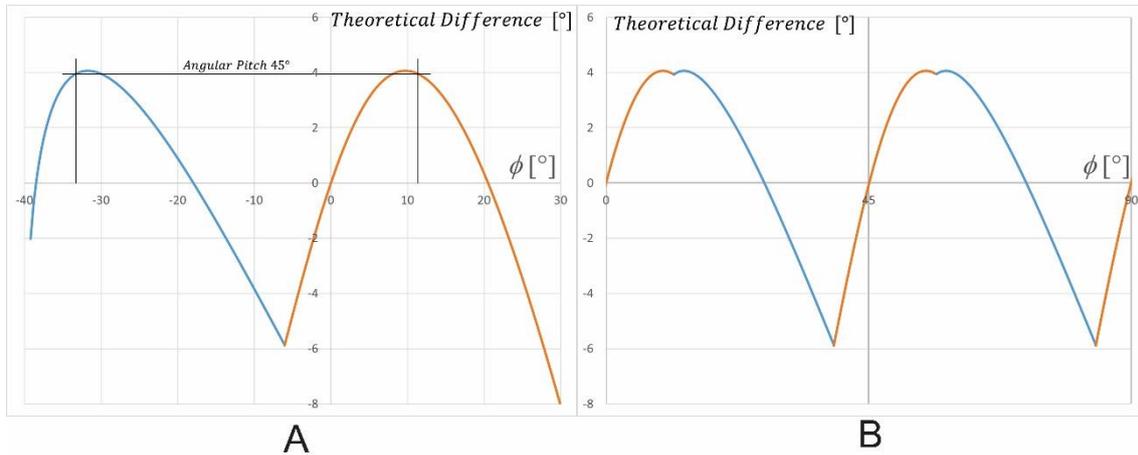

*Figure III: In Figure A, the theoretical difference is shown based on the reference system adopted for the derivations of expressions [I] and [II]. In Figure B, the reference system used in the main text is applied, resulting in the theoretical difference extended over two consecutive cycles.*

### 10- Appendix II: Optimal Separation Distance

The separation between gears is a fundamental parameter for modeling a functional mechanism. Therefore, we used computational modeling to examine how separation affects the standard deviation response of the pointer in meshed gear pairs. To do this, we input the measurements of the Antikythera Mechanism's gear pairs, as used in the main text and extracted from the supplementary material of Freeth et al. (2006). The behavior of the standard deviation as a function of the percentage separation (G) is shown in the graphs as a percentage of tooth height. Figure IV illustrates the analysis of 12 gear pairs from the mechanism, represented according to the order of magnitude of the standard deviation in a semi-logarithmic plot. The vertical axis represents the standard deviation in degrees for the meshed pair, while the horizontal axis represents the separation G as a percentage of tooth height.

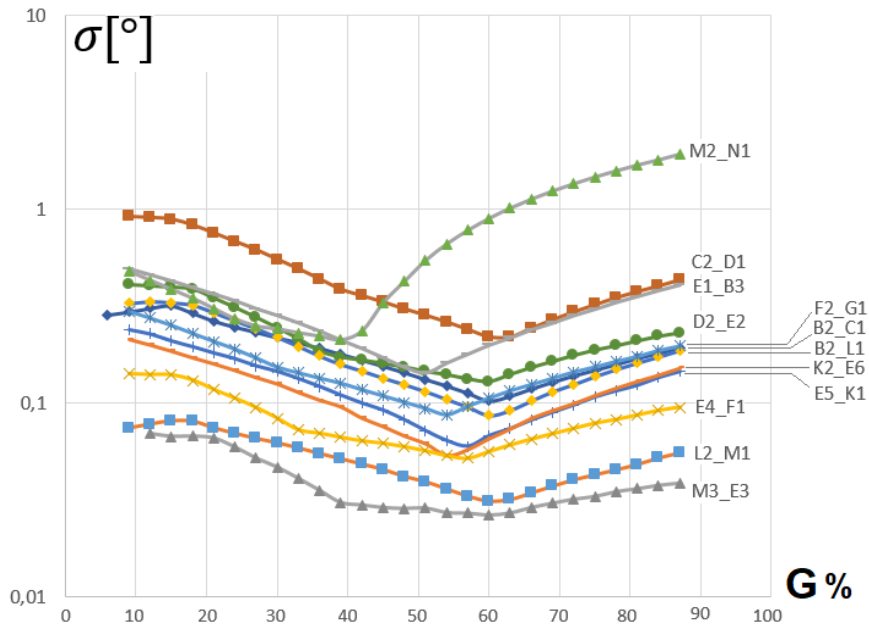

*Figure IV: Semi-logarithmic graphs showing how the separation between triangular-profile gears affects the standard deviation. The separation is expressed as a percentage of tooth height (G%). The most significant meshed gear pairs of the Antikythera Mechanism have been studied. The notation for the meshed pairs is also based on Freeth et al. (2006).*

It can be observed in the graph that there are no values below 10%. This is because the simulation found that teeth positioned too closely interfere with each other at all times. This phenomenon has a real-world counterpart in the behavior of gears when they become jammed or locked due to excessive proximity between their axes. This means that before reaching 0% separation, the teeth become jammed, as shown in Figure_V_A. There, it can be seen that one of the tips of the driving gear (I) touches one of the faces of the driven tooth[5], while at the same time, another tip (II) makes contact with the face of a preceding tooth, preventing the movement of the driven gear. Similarly, this movement restriction can also occur when the face of a driving tooth touches the tip of a driven tooth, while at the same time, the gear is unable to move due to the simultaneous contact of tip II with a face of the driving gear.

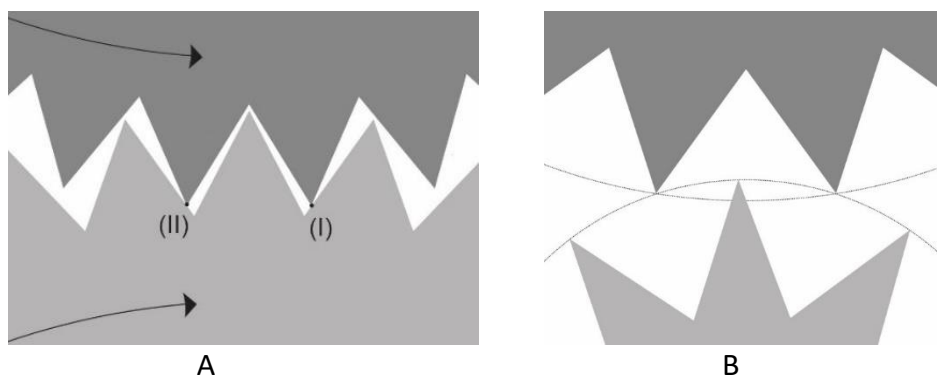

*Figure V_A and V_B: In A, a typical jamming situation is observed, where the gears are unable to move. In B, the threshold at which the gears lose contact and begin to disengage is shown.*

In the graphs of Figure_IV, there are also no values exceeding 87%. If the gears have a separation greater than this percentage, they lose contact before reaching 100% separation, and the program does not detect any tooth interference—this is what we refer to as decoupling or idle rotation. Figure_V_B illustrates a decoupling situation in which a gear fails to transmit motion to the meshed pair due to a separation below 100%. As shown in Figure_V_B, the External Radii of both gears, represented by lines, are not tangent. However, the tooth tips do not come into contact with each other, preventing the transmission of motion.

---

[5] See Figure 1 in Section 2.

In the search for a criterion to assign separation percentages to the gear pairs of the MA, the graphs exhibit a variable behavior in all cases. In general, we observe a maximum value when the separation is minimal, close to 10%. Then, a minimum occurs around 40% to 50% separation. Finally, as the separation increases to 87%, there is once again a rise in the Standard Deviation.

In the study of the MA through computational simulation, a significant correlation has been discovered between gear size and deviations from theoretical operation, measured in terms of standard deviation. Specifically, it was observed that gear pairs including at least one small gear—identified as M2-N1, C2-D1, E1-B3, D2-E2, and F2-G1—exhibit higher standard deviations in their behavior. Small gears are characterized by having a reduced number of teeth, which results in a greater proportion of tooth height relative to the total gear radius. This geometric feature means that any variation introduced by the shape of the tooth has a more pronounced impact on the interaction between gears, leading to greater variability or deviation from expected results compared to larger gears with more teeth. This finding is crucial for the design of gear pairs. However, our focus is on the overall behavior of the gear train. It is undeniable that these pairs will be critical to the functioning of the MA and will be the key elements affecting the final accuracy of the pointers.

In order to determine the valid separation values for our computational modeling of the MA and to address a broader range of possible scenarios, we have decided to consider three representative separation values: 10%, 40%, and 70%. This choice was made arbitrarily to enable the construction of MA models capable of operating Edmunds' Cases A to D. When the program was run using a triangular profile, incorporating Edmunds' errors and the three separation values, the following table was obtained.

| | Pointer | Luna | | | | | | Metonic and Saros | | | | | | Games | Exeligmos | |
|---|---|---|---|---|---|---|---|---|---|---|---|---|---|---|---|---|
| Case | G% Pairs | b2-c1 | c2-d1 | d2-e2 | e5-k1 | k2-e6 | e1-b3 | b2-l1 | l2-m1 | m3-e3 | e4-f1 | f2-g1 | m2-n1 | n2-o1 | g2-h1 | h2-i1 |
| **A** | 10 | 200/0 | 200/0 | 200/0 | 200/0 | 200/0 | 163/0 | 200/0 | 200/0 | 200/0 | 200/0 | 200/0 | 198/0 | 200/0 | 200/0 | 200/0 |
| | 40 | 0/0 | 0/0 | 0/0 | 0/0 | 0/0 | 0/0 | 0/0 | 0/0 | 0/0 | 0/0 | 2/0 | 0/0 | 0/0 | 0/0 | 2/0 |
| | 70 | 153/0 | 160/0 | 171/0 | 136/0 | 131/0 | 141/0 | 136/0 | 125/0 | 190/0 | 168/0 | 164/0 | 176/0 | 106/0 | 172/0 | 164/0 |
| **B** | 10 | 200/0 | 200/0 | 200/0 | 200/0 | 200/0 | 195/0 | 200/0 | 200/0 | 200/0 | 200/0 | 200/0 | 200/0 | 200/0 | 200/0 | 200/0 |
| | 40 | 0/0 | 1/0 | 5/0 | 0/0 | 0/0 | 0/0 | 0/0 | 1/0 | 168/0 | 99/0 | 0/0 | 5/0 | 0/0 | 1/0 | 8/0 |
| | 70 | 183/0 | 177/0 | 199/0 | 166/0 | 158/0 | 142/0 | 174/0 | 191/0 | 197/186 | 137/109 | 170/0 | 181/0 | 143/0 | 184/0 | 179/0 |
| **C** | 10 | 200/0 | 200/0 | 200/0 | 200/0 | 200/0 | 198/0 | 200/0 | 200/0 | 200/0 | 200/0 | 200/0 | 200/0 | 200/0 | 200/0 | 200/0 |
| | 40 | 5/0 | 1/0 | 197/0 | 5/0 | 10/0 | 1/0 | 6/0 | 118/0 | 200/1 | 200/0 | 7/0 | 10/0 | 10/0 | 4/0 | 17/0 |
| | 70 | 196/0 | 187/0 | 45/190 | 174/2 | 184/0 | 170/0 | 191/0 | 138/118 | 200/1 | 200/0 | 193/0 | 191/81 | 175/0 | 193/0 | 197/5 |
| **D** | 10 | 200/0 | 200/0 | 200/0 | 200/0 | 200/0 | 200/0 | 200/0 | 200/0 | 0/200 | 157/43 | 200/0 | 200/0 | 200/0 | 200/0 | 200/0 |
| | 40 | 200/0 | 195/0 | 151/49 | 185/0 | 179/0 | 148/0 | 200/0 | 68/172 | 0/200 | 105/95 | 198/0 | 200/0 | 176/0 | 200/0 | 200/0 |
| | 70 | 85/163 | 163/129 | 0/200 | 99/132 | 97/133 | 172/92 | 85/158 | 0/200 | 0/200 | 74/126 | 63/193 | 60/200 | 105/116 | 74/199 | 53/200 |

*Table I: Jamming and disengagement values of the MA gear pairs. The error values from Edmunds (Cases A, B, C, and D) were used, and the separation G% was varied at values of (10%, 40%, and 70%). The shaded values are the same as those in Table 3 of the main body of the article.*

From the table, it can be observed that the 40% separation produces the least jamming and decoupling in all cases. Therefore, this value was adopted as the criterion for all subsequent tests. It is reasonable to assume that the designers of the MA would have sought a gear arrangement that minimized jamming and decoupling as much as possible.